# Thermodynamics of the Hairpin Ribozyme from All-Atom Folding/Unfolding Simulation

Lucas G. Nivón*[§] and Eugene I. Shakhnovich[♦][§]
*Program in Biophysics and [§]Department of Chemistry and Chemical Biology
Harvard University
12 Oxford Street
Cambridge, MA 02138, USA
[♦]Corresponding Author: eugene@belok.harvard.edu

# Abstract

The structure of the self-cleaving hairpin ribozyme is well characterized, and its folding has been examined in bulk and by single-molecule fluorescence, establishing the importance of cations in the stability of the native fold. Here we describe the first all-atom folding simulations of the hairpin ribozyme, using a version of a Go potential with separate secondary and tertiary structure energetic contributions. The ratio of tertiary/secondary interaction energies serves as a proxy for non-specific cation binding: a high ratio corresponds to a high concentration, while a low one mimics low concentration. By studying the unfolding behavior of the RNA over a range of temperature and tertiary/secondary energies, a three-state phase diagram emerges, with folded, unfolded (coil) and transient folding/unfolding tertiary structure species. The thermodynamics were verified by paired folding simulations in each region of the phase diagram. The three phase behaviors correspond with experimentally observed states, so this simple model captures the essential aspect of thermodynamics in RNA folding.

# Introduction

Ribozymes perform a variety of catalytic functions in nature and the exact three-dimensional structure in the vicinity of the active site is responsible for the vast rate-enhancements affected by RNA. It is therefore of great interest to understand the exact sequence of events through which a newly synthesized RNA chain folds into the catalytically active native state before commencing cleavage, ligation, or one of many other catalytic roles. Efforts to understand RNA folding mechanism have focused largely on auto-catalytic introns, such as the Group I intron from Tetrahymena, and on smaller self-cleaving RNAs such as the hairpin, hammerhead, hepatitis delta, and Neurospora VS ribozymes (1, 2). The hairpin ribozyme was discovered in the negative strand of tobacco ringspot virus roughly twenty years ago (3, 4). It has at least two domains of secondary structure joined by tertiary interactions in the active form (5, 6). In the minimal, two-domain form, this ribozyme provides an ideal model for the formation of tertiary structure in RNA folding because it has so few interacting components. Experimental work has focused most closely on the "docking" of domains to form the tertiary structure of the active ribozyme, which is critical for catalysis (7). Single-molecule studies have measured the rates of docking and undocking for the minimal form in different cations, and even pinpointed the atoms involved in stabilization of the transition state for folding (8,9).

RNA folding has only been simulated for a pair of cases: The GCAA tetraloop RNA(10-13) and the tRNA$^{Phe}$ (14). Both have been studied using Go potentials, in which only native interactions are considered, such that atom-atom contacts in the crystal structure are attractive and contacts not present in the folded state are repulsive (15). Tetraloop folding has also been studied with physical potentials (11, 12). All-atom Go simulations have been used successfully in protein folding to predict transition state structure, even finding the conserved core residues in the folding nucleus of protein G (16, 17). Go simulations have also predicted intermediates in the folding of three-state folders barnase, RNase H and CheY (18). Simulations combining phi-value constraints with Go potentials have verified and refined models of the TSE and identified the most critical residues in the folding nucleus (19-21). In the case of nucleic acids, Go simulations predict the correct "zipping" pathway for the a GCAA tetraloop hairpin RNA (10), and a combined Go/AMBER94 simulation of tRNA predicted intermediates along the folding path in qualitative agreement with experimental work by Sosnick (14, 22).

The minimal hairpin ribozyme (two-way junction, 2WJ) has fewer domains than tRNA, and the thermodynamics and kinetics of tertiary structure formation have been well characterized for both the minimal and full (four way junction, 4WJ) forms of the RNA. In addition, the transition state for tertiary structure formation has been analyzed in some detail using single-molecule techniques (9). Theoretical study of the ribozyme has used the nonlinear Poisson-Boltzmann equation to estimate the magnesium uptake of various hypothesized transition state structures (9). With the advent of techniques for the simulation of larger RNAs, it is now of great interest to simulate the full folding trajectory of the hairpin ribozyme at all-atom resolution using a Go potential.

Metallic cations play a vital role in the formation of RNA tertiary structure. Tertiary structure necessarily involves placing negatively charged backbone phosphates in close proximity with one another. Monovalent cations can mitigate the repulsion between negative charges, allowing folding to proceed, but divalent cations do so more efficiently and appear to be necessary for full formation of tertiary structure in many, but not all, cases (23, 24). Early work by Crothers and co-workers on tRNA demonstrated the existence of three major phases as a function of magnesium and temperature (25). In this phase diagram the RNA shows native (folded) structure at high cation concentration and low temperature, coil at high temperature (completely unfolded) and "extended forms" at low temperature and cation concentration. More recently, similar results have been observed for the hairpin ribozyme using fluorescence techniques, demonstrating a dynamic phase (modeled as docking/undocking) at low magnesium, entirely unfolded structures at high temperature (26), and a stably docked state at higher magnesium and low temperature (9).

The interaction of magnesium with RNA has been modeled as a combination of specifically bound ions along with a diffuse cloud surrounding the RNA, both of which act to preferentially stabilized compact RNA structures at higher cation concentrations (27). This framework allows a detailed description of the influence of $Mg^{2+}$ on the stability of various RNAs in unfolded, intermediate and native states. Such models of the role of $Mg^{2+}$ in RNA thermodynamics have not been extended to the study of RNA dynamics or folding, but rather have focused on predicting the stability and concentration of bound ions for static structures. In the simulation of the tRNA$^{Phe}$ the effects of $Mg^{2+}$ were modeled in a fashion very similar to the present work, separating contacts into secondary or tertiary categories (14). Here we extend that analysis by studying the influence of varying cation concentration within the Go model.

We develop a modified Go potential with a simple accounting for the effects of cation concentration. This model allows a full consideration of the thermodynamics of RNA folding at varying $Mg^{2+}$ or $Na^+$ concentration, treating all interactions as non-specific. It also results in full folding simulations of the ribozyme which verify the thermodynamic results of unfolding simulations. These atomically detailed folding simulations also allow us to observe the tertiary interactions responsible for folding as they are formed, and we plan to model the TSE for docking using extensive $p_{fold}$ analysis in future work. The thermodynamic results predict three major regions of cation/temperature phase space, and make quantitative predictions about the boundaries between different regimes.

## Results

### Model and Unfolding Simulations at τ/S = 0

The present model uses a modified Go potential, where interactions are categorized as tertiary or secondary based on the native structure (See Model and Methods for complete description). The strength of different interactions can be scaled by separate tertiary structure interaction strength (τ) and secondary structure strength (S) parameters. For example at τ = 0 and S = 1, only secondary interactions contribute to the energy, while tertiary contacts between domains are ignored. Since cations preferentially stabilize tertiary structure, varying the τ/S ratio corresponds to adjusting the ion concentration (25, 28, 29). In this model we hold S = 1 and only vary τ, to mimic adjustments in ionic strength of tertiary interactions. This model only captures non-specific binding effects, and does not explicitly model ions bound at specific sites. The model allows us to systematically study the behavior of the ribozyme as a function of τ/S and temperature, and compare these results with measured thermodynamic data. Furthermore it provides an independent method to estimate the relative strength of tertiary and secondary contacts in the ribozyme, as will be examined below and in the Discussion.

A model of the two-domain minimal hairpin ribozyme, here called hp2d, was constructed from the crystal structure of the full ribozyme (Figure 1; see Model and Methods). In all subsequent results, the structure hp2d will simply be referred to as "the hairpin ribozyme". To evaluate the model, unfolding simulations were performed with $\tau/S = 0$ at a variety of temperatures. At low temperatures undocking was observed without substantial dissolution of secondary structure in domain A or B (Figure 2A, B). Undocking here refers to the loss of tertiary structure (Domain A – Domain B interaction) without complete unzipping of either secondary structure domain.

Unfolding via undocking occurs to an intermediate ensemble of states which consist mostly of a variety of undocked structures without any A-B stacking (Figure 2C) but with intact secondary structure within domains. Structures with stacking between A and B, where the helix continues (with an interruption) via stacking between the ends of domain A and B, also occur (Figure 2D) with some stability. Brief re-docking events are also observed, for example at ~15,500 in Figure 2A, to a structure very close to the native (Figure 2E shows an overlay of the re-docked structure (blue) with the native hp2d (orange)).

**Phase Diagram of Hairpin Unfolding**

Cation concentration and temperature define a phase diagram for the hairpin ribozyme. At high temperature, one expects all molecules to unfold. At low temperature and high ionic strength a stably docked structure should be observed. From *a priori* considerations it is impossible, however, to make quantitative predictions about the nature of this phase diagram (e.g., where boundaries will lie), or to deduce whether phases other than folded/unfolded will occur. To map out this space, we conducted unfolding simulations of the hairpin ribozyme over a range of temperatures (all temperatures are Monte Carlo temperature) from 1 to 10 and $\tau/S$ ratios of 0 to 1.0 (Figure 3).

Comparison of overall structure formation (Figure 3A (left)) with secondary structure alone (Figure 3A (right)) shows at least three distinct regions of the phase diagram We track the total dRMS at each point in the phase diagram in Figure 3A, left – this quantity indicates both secondary and tertiary structure formation (low dRMS; light in the figure) or dissolution (high dRMS; dark in the figure). In Figure 3A (right) the sum of the dRMS values for each of domain A and B alone is plotted – this indicates secondary structure formation/dissolution without tertiary structure influence. By comparing overall structure (left) with secondary structure alone (right) a discrepancy is apparent in the lower-left hand corner (low $\tau/S$ and T), where secondary structure is intact but total structure is not (dRMS is non-zero).

At high temperature, the hairpin unfolds, with high total and intra-domain dRMS. At low temperature and high $\tau/S$ the structure is native, with low dRMS for both the entire structure and within each isolated domain. Interestingly, at low temperature and low $\tau/S$, a third phase emerges with intermediate total dRMS but very low intra-domain dRMS. In other words, secondary structure is entirely intact, while the overall structure has non-native dRMS. In trajectories in this region undocking and docking are observed. Note that domains A and B in isolation unfold at T ~ 4.5, with only a very slight preferential stability at lower $\tau/S$. Indeed, for low $\tau/S$ and low T secondary structure is nearly entirely complete, while tertiary structure is partly dissolved.

To more clearly illustrate the three phases we draw a schematic of the unfolded, folded and docking regions in Figure 3B. In the schematic solid lines indicate the midpoint of a transition between two states, while dashed lines indicate the beginning/end points of a transition. It is interesting to note that docking/undocking behavior ranges over $\tau/S$ from 0 up to slightly more than 0.3, which gives an estimate for the ratio of tertiary interaction strength to secondary strength in the docking region. $\tau$ represents an average over all tertiary interactions, including hydrogen bonds, but also stacking interactions or salt-bridges, and S likewise averages over all interactions within the domains. This phase diagram suggests that, averaged over all contacts, tertiary interactions contribute one third the enthalpy of secondary interactions. We discuss the implications of this result in the discussion.

**Hairpin Folding Trajectories**

To verify the equilibrium behavior observed over the phase space of $\tau/S$ and temperature via unfolding simulations of the hairpin ribozyme, it is necessary to also find the same results for folding simulations. This is of some computational difficulty, because while unfolding occurs rapidly, folding, even with a Go potential, can take much more computational time. To reduce the size of the computational load, we focus on a set of three representative sections in phase-space: (1) in the unfolded region, (2) in the docking region and (3) in the folded region (Figure 3B; see Methods for details). In each region we run simulations from the undocked structure long enough to observe a significant number of folding events.

All three regions of the phase diagram observed by unfolding simulation can be recapitulated in folding experiments. In region 1 no folding is observed during the 300,000 step runs. The simulations explore regions of structural space with both secondary and tertiary structure dissolved, which can be seen clearly by tracking $Q_{sec}$ and $Q_{tert}$, the fractions of native secondary and tertiary contacts, respectively (Figure 4). In region 2 docking/folding was seen in 47/104 (45%) cases (where folding is defined as attainment of a structure with dRMS less than 0.5 Angstroms by 300,000 steps), confirming that folding is possible, and that docking (to low dRMS and high $Q_{tert}$) occurs. In this region, secondary structure is largely intact (high $Q_{sec}$), and the unfolded ensemble makes many forays into nearly-docked structures (Figure 5) and occasionally docks but does not remain docked (Figure 5). Finally, in region 3 41/116 (35%) structures fold (dock) within the allotted simulation time of 300,000 MC steps -- it appears that docking occurs less frequently in region 3 than 2. Prior to folding in region 3, the RNA explores the unfolded ensemble with sampling in the vicinity of the docked structure, and after folding the structure does not undock/unfold (Figure 6) – it remains stable for the duration of simulations.

To better quantify folding in regions 2 and 3, docking/undocking and docked, respectively, we measured the dwell time of simulations in the docked structure in both regions. A docked structure here is defined as attainment of a structure with dRMS to the native of less than 0.5 Angstroms. In region 2 docking dwell times are exponentially distributed (Figure 7A) as expected for a single-transition-state undocking transition. In contrast, in region 3 docking is irreversible on the time scale of the simulations, so dwell times in the docked state (measured from folding until the end of the trajectory) are very widely distributed, far from a single exponential (Figure 7B). The kinetic behavior indicated by the docking dwell time distributions is consistent with dynamic docking in region 2 and stable docking in region 3.

## Discussion

We have constructed a simple simulated model for the role of cations in RNA folding and have presented a detailed phase diagram for a functional ribozyme, the hairpin ribozyme derived from Tobacco Ringspot Mosaic satellite RNA. We have also shown the first folding simulations of the hairpin ribozyme to the native structure. The model predicts a three-state phase diagram for the ribozyme, with folded (docked), unfolded and docking/undocking regions. This behavior may be very general for structured RNAs. A three or four-state phase diagram for a RNA was first observed by Crothers and co-workers in the 1970's for tRNA (25). The three-state phase diagram for tRNA in varying magnesium concentrations was confirmed by NMR exchange measurements (30). More recent experiments with the hairpin ribozyme suggest a three-state phase diagram for the minimal form (the same form studied here) (9, 26). The folding simulations from the undocked state recapitulate these phase regions and offer trajectories of tertiary structure formation at atomic-level resolution.

The model presented here for the role of cations in folding is nearly identical to that used by Sorin and co-workers (14) for the tRNA$^{Phe}$, but here we present a systematic analysis of the role of the cation in folding, which was not carried out for the tRNA. This model is based on the assumption that magnesium or sodium ions have little effect on RNA secondary structure while strongly stabilizing tertiary structure (27). It also presumes that all tertiary interactions are influenced by cations. This may be a poor assumption for certain interactions, for example hydrogen bonds which may not be context-dependent, such as the G+1:C base pair between Domains A and B in the hairpin ribozyme (5). However, it serves as a useful first-order approximation to describe the majority of tertiary interactions.

Fluorescence experiments in bulk and at the single-molecule level suggest a three-state phase diagram for the 2WJ version of the hairpin ribozyme, the same "minimal" form studied here. Klostermeier and Millar (26) carefully dissected secondary and tertiary structure transitions for the 2WJ and 4WJ (which includes the two naturally occurring C and D domains in addition to A and B). They employed simple UV absorbance to measure secondary structure transitions and bulk FRET (with the donor and acceptor attached at the ends of domains A and B, respectively) to measure tertiary interactions between domains. For the 2WJ they found a single transition in secondary structure as a function of temperature, with a melting temperature of 321° K. In our model secondary structure melts in a single transition at all τ/S (as measured by dRMS within domains A and B; Figure 3A, right), mimicking the single transition of the 2WJ observed by UV absorbance.

A global fit to a two-state docking model for tertiary structure formation revealed a transition between docked/undocked at temperatures below 315° K , to completely unfolded above 315° K . This transition corresponds to the transition between region 2 (docking/undocking) and region 1 (unfolded) in the simulation (Figure 3B). Single-molecule work reveals the region 2 – region 3 (folded) transition. Zhuang and co-workers measured the rates (and therefore stability) of the docking/undocking transition as a function of magnesium, and observed a transition to mostly docked ribozymes at high magnesium (above ~100 mM) or very high sodium concentration (above ~1M) (9). Together these observations demonstrate the existence of a docking/undocking – unfolded transition and a docking/undocking – folded transition. The predicted single-step transition between stably folded (region 3) and unfolded (region 1) at high cation concentration (figure 3B) has not been directly measured for the hairpin ribozyme. Our results suggest that the ribozyme should melt in a concerted single step (combining tertiary undocking and secondary unfolding) at high cation concentrations. This behavior is in fact observed for tRNA: At lower magnesium levels, tertiary structure melts first, followed by various secondary structure elements, while at higher magnesium levels a single melting step is observed at around 60° C (30). The three-state phase diagram of structured RNAs may be relatively generic, occurring in all RNAs with tertiary structure, or at least in those RNAs with a single tightly-interconnected region of tertiary interactions.

The thermodynamic results presented here make a prediction about the ratio of tertiary to secondary interaction energies in biochemically and physiologically relevant conditions. We find that $\tau/S$ is greater than 0 but smaller than 0.3 in the "docking" region of the phase diagram, which corresponds to an intermediate range of cations, for example less than ~100 mM magnesium (9). In our model $\tau$ is identical for all tertiary interactions – it acts effectively as the mean tertiary interaction, including any stacking or hydrophobic interactions, hydrogen bonds and ionic interactions. Similarly, S averages over all secondary interactions – mostly stacking and prototypical Watson-Crick base pairing. It is important to note that hydrogen bond strengths are context-dependent (31), and that other atom-atom tertiary interactions may depend on ionic strength for their stability (or extent of repulsion, in the case of coulombic interactions).

We can independently estimate an expected value of $\tau/S$ for hydrogen-bonding terms alone based on measurements of context-dependent hydrogen-bond strength in other RNA systems and from measurements in the hairpin ribozyme 2WJ and 4WJ. The work of SantaLucia and co-workers gives a set of measurements of the strength of RNA hydrogen bonds, separating canonical Watson-Crick (WC) hydrogen bonds from bonds in an RNA turn (measurements in 100mM NaCl) (31). The stem hydrogen bond has a $\Delta\Delta G$ of 5.3 kJ/mol upon mutation. The average over all intra-loop hydrogen bonds (non-stem bonds) gives $\Delta\Delta G = 1.8$ kJ/mol, so the loop hydrogen bonds are ~33% as strong as intra-stem (WC) hydrogen bonds. The analogy between tertiary hydrogen bonds in the hairpin ribozyme and intra-loop bonds in an RNA turn is a loose one, but this estimate shows that context-dependence could easily result in a $\tau/S \sim 0.3$.

For a more realistic estimate we turn to mutational studies of the hairpin ribozyme by Klostermeier and co-workers (32), where energetic contributions of some of the tertiary hydrogen bonds were measured at 12 mM $Mg^{2+}$. The ribozyme has three main motifs of tertiary interactions between domains A and B: U42 packing, the ribose zipper (involving bases A10, G11, A24 and C25) and the G+1: C25 base pair. The energetic contributions of the ribose zipper hydrogen bonds were measured, with a mean $\Delta\Delta G$ of 2.8 kJ/mol (another measurement in the Tetrahymena group I intron P4-P6 domain ribose zipper puts the range at 1.7-2.0 kJ/mol (33)). In that work other interactions were not measured for the 2WJ. Using this estimate and the SantaLucia measurement of a stem hydrogen bond, 5.3 kJ/mol, we estimate the mean tertiary hydrogen bond to be 53% as strong as a secondary hydrogen bond. According to our simulations, $\tau/S$ is in the range of 30%, which would predict that the $\Delta\Delta G$ of other tertiary interactions (other hydrogen bonds, non-hydrogen bonding terms, and repulsive terms) are significantly lower, in order to arrive at a $\tau/S$ less than 53% overall. The energetic contribution of one hydrogen bond involved in U42 packing was measured to be $\Delta\Delta G = 2.4$ kJ/mol in a single-molecule mutation study (9), which is not sufficiently low to explain a $\tau/S \sim 30\%$. Our results suggest that a full examination of both hydrogen bonding and other tertiary interactions in the context of the hairpin ribozyme would reveal $\tau/S$ to be less than or near 30% at intermediate magnesium or sodium concentrations.

The model presented here allows for the construction of a phase diagram for RNA folding of the hairpin ribozyme. Remarkably, given the simplicity of the model, it predicts the existence of an intermediate region on the phase diagram corresponding to formed but docking/undocking helixes, and makes predictions about the character of transitions between the three different behaviors as a function of temperature and ionic strength. All of the predicted transitions in this phase diagram have been observed for the hairpin ribozyme or for tRNA. The model also predicts the ratio of tertiary to secondary mean interaction strengths in a way that does not involve serial mutations and measurements of ΔΔG. These are the first atomic-level simulations of the folding of the hairpin ribozyme, and will allow for predictions of the kinetics (Transition state, cation dependence) of hairpin docking in future work.

## Model & Methods

Our model closely parallels the all-atom protein model of Shimada *et al.* in overall concept, and is a modified version of the RNA model of Nivón and Shakhnovich (10). The latter model was modified to account for the effects of differential tertiary/secondary interaction strength at varying cation concentrations.

Based on the secondary structure diagram of the RNA in question, the user inputs the boundaries of secondary structures (for the hairpin construct here, domain A constitutes C1 through A33; domain B A34 through A70). During the initial read-in of a structure for folding, atom-atom contacts are categorized as secondary (domain i – domain i) or tertiary (domain i – domain j). In all subsequent energy calculations, the total number of secondary and tertiary atom-atom contacts are counted separately. The energy of a conformation ($E_{Go}$) is then calculated as:

$$E_{Go} = S * n_s + \tau * n_t \qquad (1)$$

Where $n_s$ and $n_t$ are the number of native secondary and tertiary contacts respectively. S is the secondary structure strength parameter, and $\tau$ is the tertiary structure parameter. Here we fix S=1 and only vary $\tau$.

At $\tau/S = 0$, no tertiary interactions are stabilized, simulating low cationic concentrations. Higher ionic strength is modeled by $\tau/S > 0$. Specific cation binding sites are not explicitly modeled, so this construction can only model non-specific interactions of cations, such as magnesium or sodium, with the ribozyme.

### Construction of the Hairpin Ribozyme Model

The structure of the four-way hairpin ribozyme was taken from the crystal structure, PDB ID 1M5K (Figure 1A, B) (5). In this structure, a U1A binding loop was attached at the end of Domain B for ease of crystallization. Note that in the figure only one of the two RNAs in the unit cell is shown, and the U1A protein has been removed. For the simulation, the U1A loop was removed and replaced with the loop structure of loop B from NMR studies (PDB ID 1B36) (34).

The wild-type ribozyme has four domains A-D (Figure 1), but the active site resides in domains A and B, and a construct with C and D cut out is catalytically functional (7). We simulated the minimal Domain A- Domain B ribozyme which has been studied both in bulk and at the single molecule level (7, 8) so that the *in silico* predictions would be most directly comparable with experiment. Therefore domains C and D were removed, and the substrate and ribozyme strands (at the end of Domain A) were joined by an inserted loop (loop B from 1B36). The final schematic is shown in Figure 1D, with the structure in 1C. This structure of the two-domain structure will be abbreviated as hp2d to differentiate it from the original four-way junction crystal structure.

### Hairpin Unfolding simulations

To evaluate the simulation methodology, unfolding runs at $\tau = 0$, $S = 1$ were performed for 20,000 MC steps. dRMS, along with the dRMS of sub-regions of domains A and B, were recorded every 100 steps. The total dRMS reports on the combined secondary/tertiary structure of the ribozyme, while the dRMS of an individual domain reports on the secondary structure of that domain alone.

**Phase Diagram of Hairpin Unfolding**

Unfolding simulations were carried out with $\tau/S$ parameter (modeling cation concentration) ranging from 0 to 1 (steps of 0.1) and T varying from 1 to 10 (steps of 1.0), for a total of 110 points in the $[\tau/S]$ / T phase diagram. Simulations were as described for preliminary unfolding simulations (above). dRMS and domain dRMS were recorded at all points. dRMS over the entire molecule contains contributions both from tertiary and secondary structure, since both undocking and helix-unfolding will cause deviations from low dRMS. The intra-domain dRMS is calculated separately for four sub-regions of structure within the hairpin, the upper and lower halves of domains A and B. Each intra-domain dRMS is not effected by global changes in structure (such as undocking) but instead only reports on the extent to which that domain is native-like Three runs were conducted at each $[\tau/S]$/T combination, for a total of 330 independent unfolding runs, and the average result calculated for the various observables at each point.

**Hairpin Folding Simulations from Undocked States**

Structures in unfolding simulations with high overall dRMS but native-like secondary structure were isolated and the one with most intact helices (UD) was selected for subsequent folding simulation. Folding simulations, with the Go potential defined by hp2d, were carried out for 300,000 MC steps starting from UD. Three regions in the phase space were chosen for study: (1) unfolded region [T = 8, $\tau/S = 0.2$], (2) docking/undocking region [$\tau/S = 0.2$, T = 2], (3) folded region [$\tau/S = 0.9$, T = 2]. Other simulations were also conducted at $\tau/S = 0.2$, T = 3,4 and $\tau/S = 0.3$, T = 2,3,4, but most results reported are in the three regions described above.

## Acknowledgements

We thank Professor Xiaowei Zhuang and Greg Bokinsky for discussions about the hairpin ribozyme and the cation dependence of docking/undocking. L.G.N. is supported by a Graduate Fellowship from the Fannie and John Hertz Foundation.

Mechanism of Thermal Unfolding of Escherichia coli Formylmethionine Transfer RNA. *J. Mol. Biol.* 87:63-88.

## Figure Legends

**Figure 1**. Structure of the four-way junction hairpin ribozyme and the two-way junction model used for simulation. The crystal structure, 1M5K, (absent U1A co-crystallization protein) is shown with bases in blue, backbone in grey (A) (5). A schematic diagram of the secondary structure of the ribozyme (B). The atomic structure of the model of the two-way junction, hp2d, with loops modeled in from the structure of the isolated Domain B NMR structure(1B36) (C) (34). A schematic secondary structure diagram of hp2d.

**Figure 2**. Sample trajectories and structures from unfolding simulations of hp2d with $\tau/S = 0$. Trajectories of total dRMS, region dRMS and fraction of native contacts as a function of number of MC steps show undocking with brief re-docking (A) and undocking without re-docking, but with excursions to an extended "stacked" structure (B). For A and B total dRMS (red) and the sum of domain A + domain B dRMS (green) is shown in the top panel. The fraction of native secondary contacts ($Q_{sec}$; dark blue) and native tertiary contacts ($Q_{tert}$; light blue) is shown in the bottom panel. Undocked structures occupy a wide swath of structural space away from the stacked structure (C). An example of the stacked structure (D). An overlay of a re-docked structure (blue) and the native structure (orange) shows that the re-docked structure is native (E)

**Figure 3**. $\tau/S$ and temperature phase diagram of the hairpin ribozyme. The phase diagram was mapped out over $\tau/S = 0.0$ to $1.0$ (steps of 0.1) and $T = 1$ to $10$ (steps of 1). Three folding simulations were carried out at each spot in the phase space. The averaged total dRMS at each spot, indicating overall structure formation, is shown in (A, left). The averaged dRMS for domain A plus domain B, indicating secondary structure alone, is shown in (A, right). Higher dRMS (less native) is darker. Comparison of the overall dRMS with the isolated region dRMS shows a discrepancy at low T and $\tau/S$, where overall dRMS is intermediate while regional dRMS (right) is very low, indicating loss of tertiary structure with maintenance of secondary structure.

By superimposing overall dRMS (left) and domain A + B dRMS (right), one can delineate the three state phase diagram. The regions of phase diagram are shown schematically based on the results of (A) in the diagram (B), showing folded, unfolded and docking regions.

**Figure 4**. Folding simulations from the undocked structure, in region 1, unfolded. A structure with intact secondary domains, UD, was extracted from unfolding simulations. Folding runs were carried out in three regions of phase space: (1) unfolded (this figure), (2) docking/undocking (Figure 5) and (3) folded without subsequent undocking in region 4 (Figure 6). Total dRMS (red) and the sum of domain A + domain B dRMS (green) is shown in the top panel. The fraction of native secondary contacts ($Q_{sec}$; dark blue) and native tertiary contacts ($Q_{tert}$; light blue) is shown in the bottom panel. Note that overall and domain dRMS quickly jump to high values and that $Q_{sec}$ rapidly equilibrates to a low value, indicating unfolding of domains, while $Q_{tert}$ never rises, indicating a lack of any docking.

**Figure 5**. A folding trajectory in region 2 with docking and subsequent undocking. Note in the upper trace that the domain dRMS remains low, while overall dRMS only falls to low values near 230,000 steps, indicating folding. The $Q_{sec}$ remains high throughout, and $Q_{tert}$ only rises during docking.

**Figure 6**. A folding trajectory in region 3 with docking but no subsequent undocking. Undocking is never observed in region 3. The upper trace indicates folding to low overall dRMS around 120,000 steps. The fraction of native tertiary contacts jumps to a stable high value upon folding, while $Q_{sec}$ is not affected.

**Figure 7**. The dwell time distribution for docked structures (dRMS < 0.5 Angstroms) in region 2 (A) and 3 (B). Dwell times were determined from folding simulations starting from the undocked hairpin structure, with 104 trajectories in region 2 and 116 in region 3. The dwell times in region 2 are exponentially distributed (A) as expected for a two-state undocking reaction coordinate. In region 3, the dwell times are not exponentially distributed (B), as expected for an irreversible folding reaction.

A

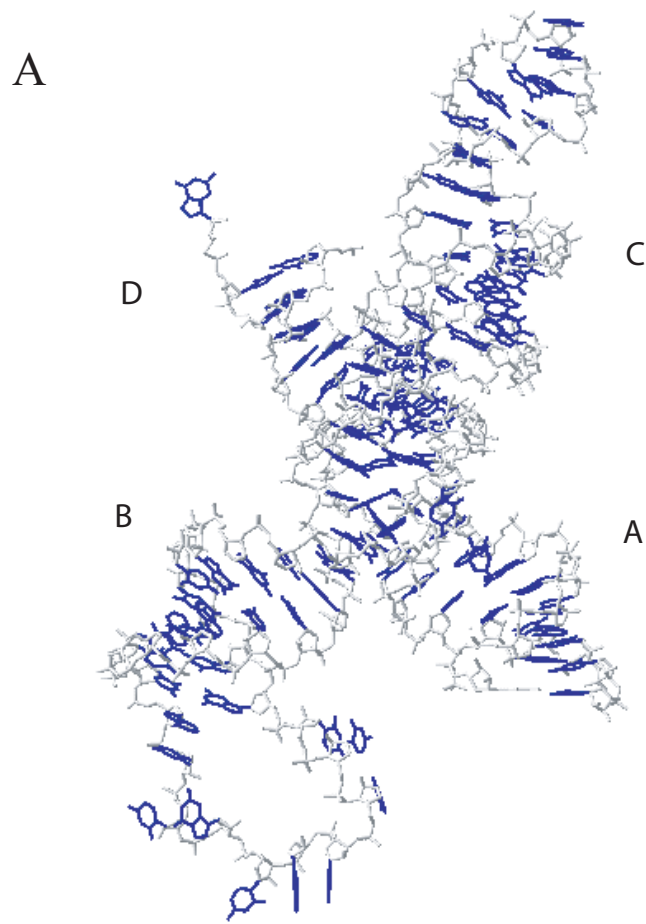

B

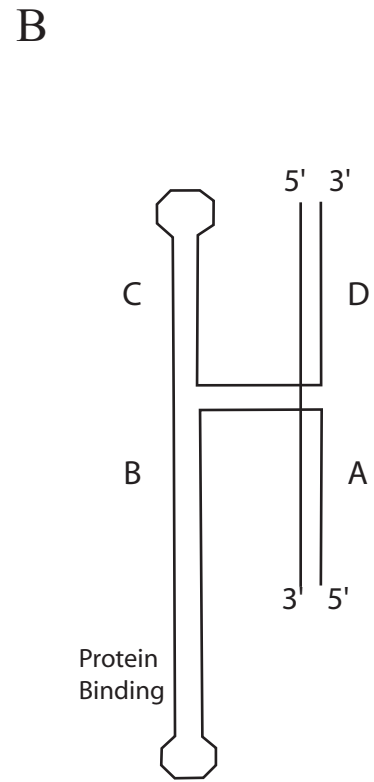

C

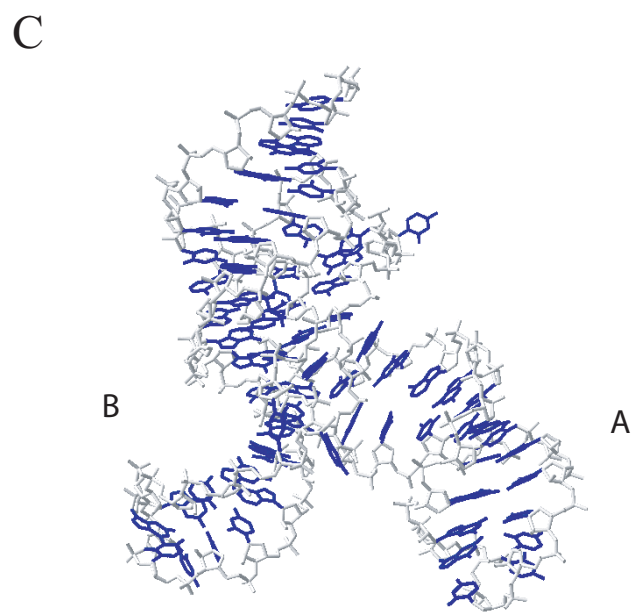

D

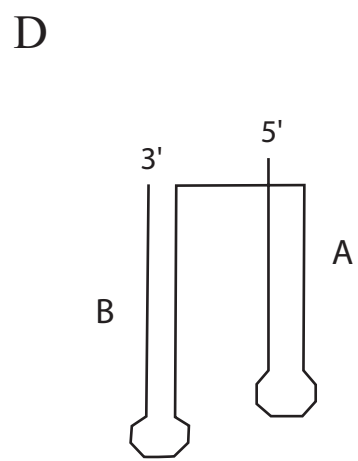

Figure 1

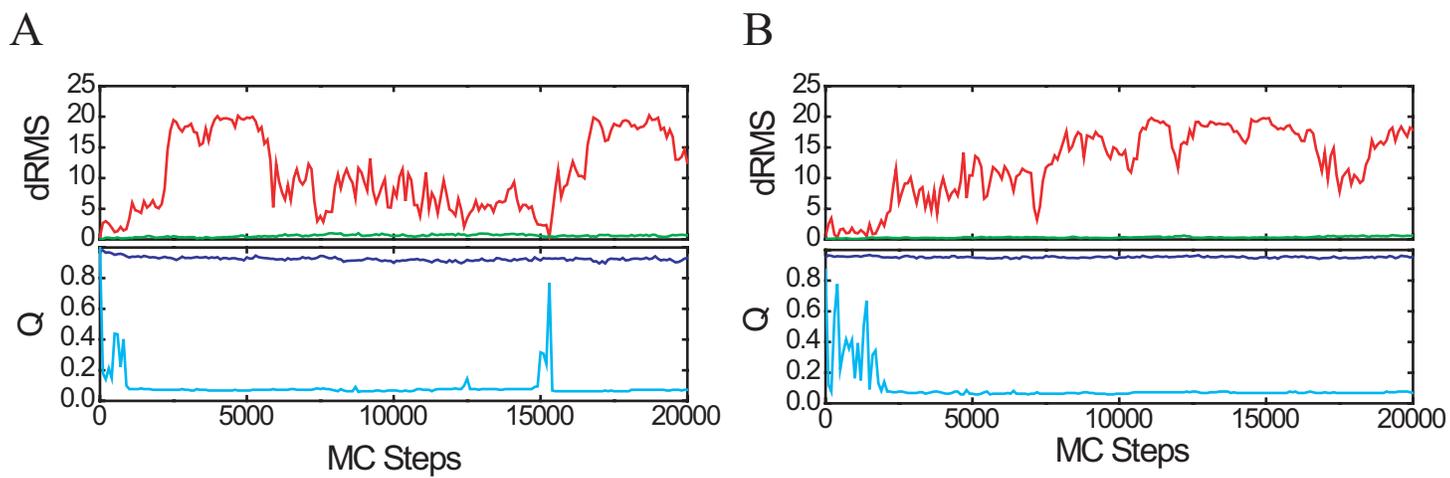

C

D

E

Figure 2

A

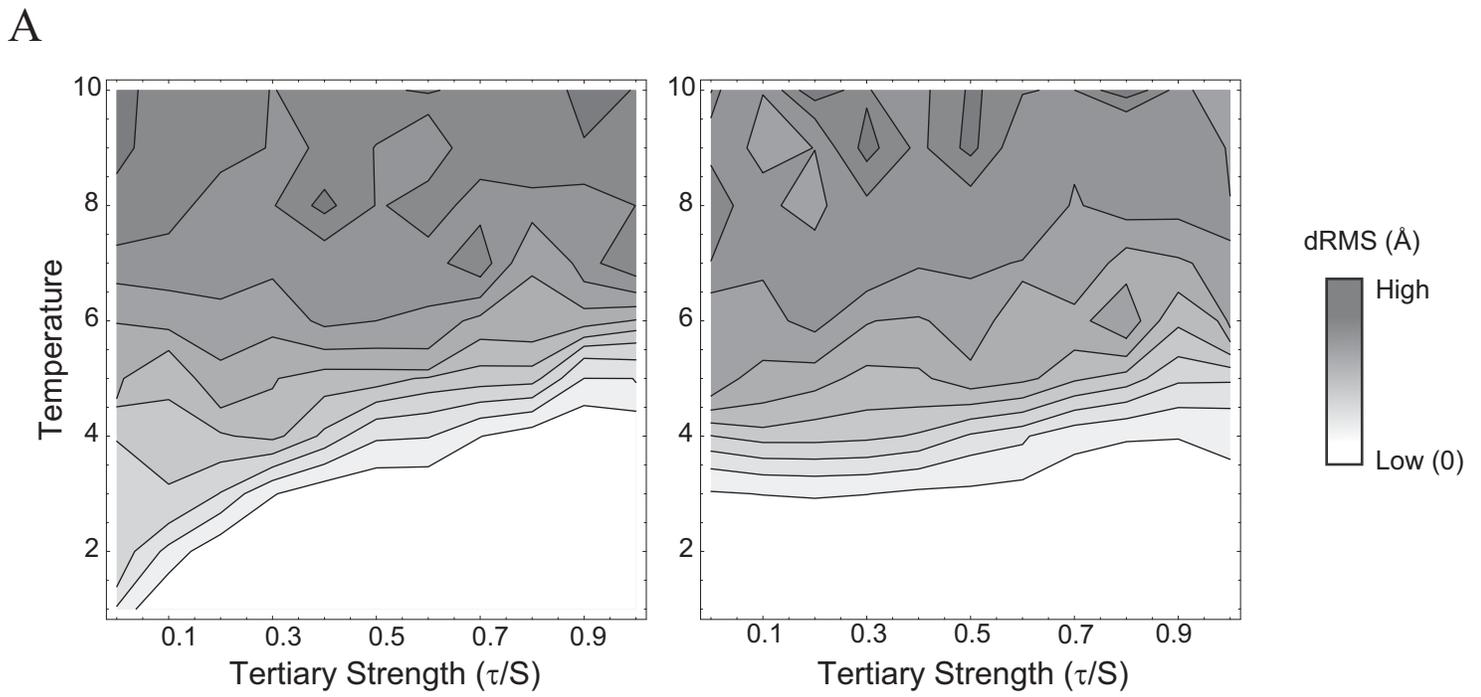

B

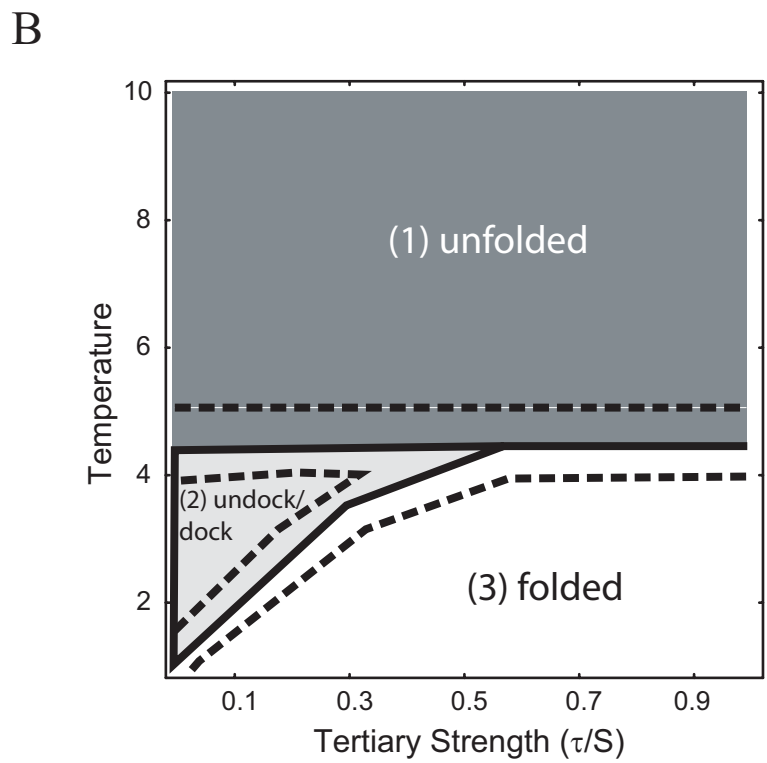

Figure 3

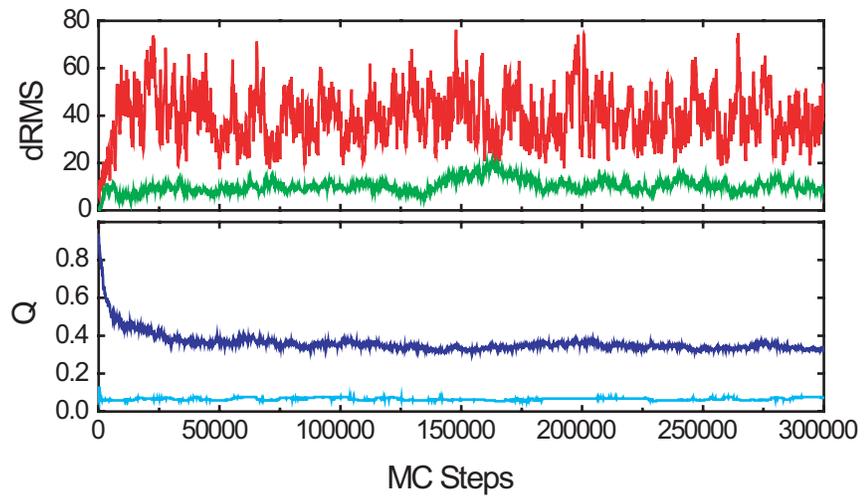

Figure 4

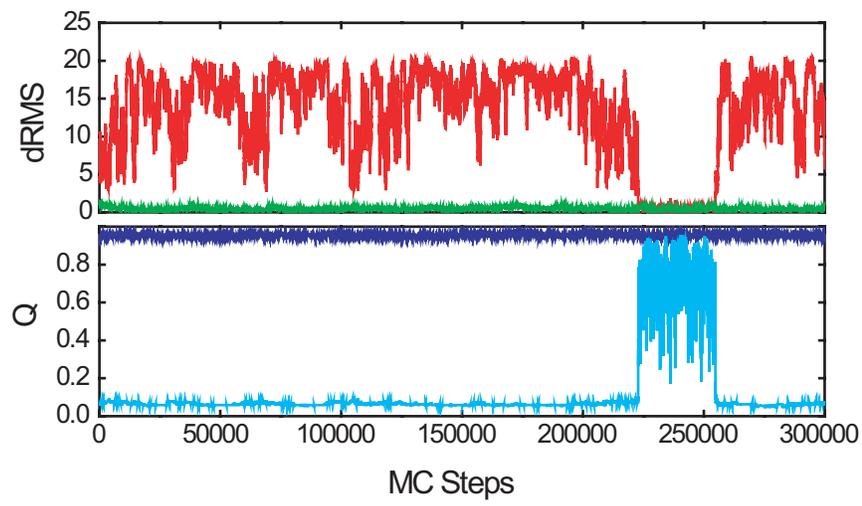

Figure 5

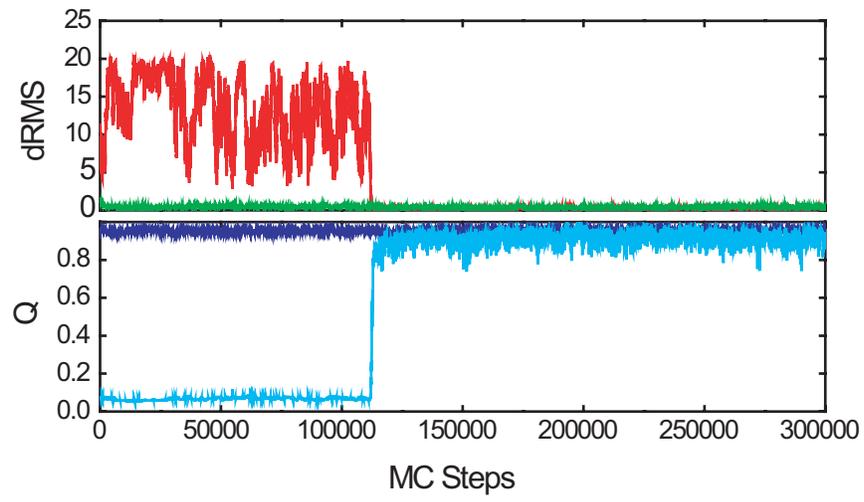

Figure 6

A

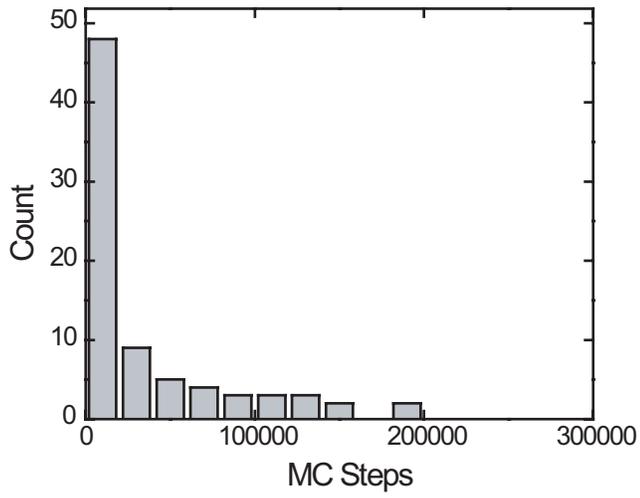

B

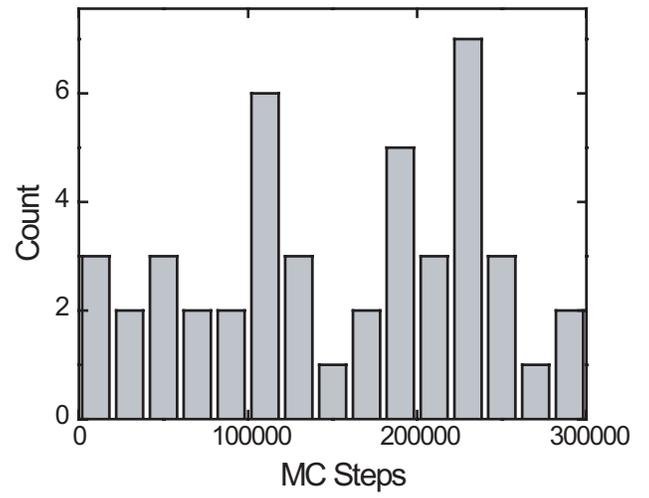

Figure 7